\documentstyle[aps,amssymb]{revtex}   
\newcommand{\be}{\begin{equation}}    
\newcommand{\ee}{\end{equation}}
\newcommand{\ba}{\begin{eqnarray}} 
\newcommand{\ea}{\end{eqnarray}}

\title{ Large Deviation Approach to the Randomly Forced Navier-Stokes 
Equation}
\author{R.~Collina$^{1,2}$, R.~Livi$^{3,4}$ and A.~Mazzino$^{1,2,5}$}
\address{
$^1$ Dipartimento di Fisica, Universit\`a di Genova, Via Dodecaneso 
33, I--16146 Genova, Italy\\
$^2$ Istituto Nazionale di Fisica Nucleare, Sez. di Genova, Via Dodecaneso
33, I--16146 Genova, Italy,\\
$^3$ Dipartimento di Fisica, Universit\`a di Firenze, Via Sansone 1,
I--50019 Firenze, Italy\\
$^4$ Istituto Nazionale di Fisica della Materia, UdR di Firenze e Istituto 
Nazionale di Fisica Nucleare, Sez. di Firenze, Via Sansone 1,
I--50019 Firenze, Italy\\
$^5$ Istituto Nazionale di Fisica della Materia, UdR di Genova, Via Dodecaneso 
33, 
I--16146 Genova, Italy\\}
\date{\today}            
\begin{document}
\draft
\maketitle
\begin{abstract} 
The random forced Navier-Stokes equation can be obtained as a variational
problem of a proper action. By virtue of incompressibility,
the integration over transverse components of the fields allows to cast
the action in the form of a large deviation functional.
Since the hydrodynamic operator is nonlinear, the functional integral
yielding the statistics of fluctuations can be practically computed by
linearizing around a physical solution of the
hydrodynamic equation. We show that this procedure yields
the dimensional scaling predicted by K41 theory at the lowest
perturbative order, where the perturbation parameter is the inverse
Reynolds number.  Moreover, an explicit expression of the prefactor of
the scaling law is obtained.
\end{abstract} 
\section{Introduction}
A field theoretic approach to the study of the random stirred Navier-Stokes
equation (rsNSE) can be traced back to the seminal paper by Martin, Siggia and
Rose \cite{MSR}. This was the starting point for the application of many
field-theoretic strategies, e.g. diagramatic expansions, renormalization
group methods \cite{ren}  
(for recent developments and applications the reader can be addressed to
\cite{Dev}), instanton-based approaches (for 
applications of instantonic methods in turbulence see, e.g., 
\cite{ist1,ist2,ist3,ist4} and references therein) and combinations of 
them \cite{GIL}. 
The many technical difficulties encountered in
developing these approaches avoided to gather conclusive achievements.\\
In this paper we show that one step
forward along this field-theoretic approach allows one to cast the
action associated with the rsNSE into the form
of a large deviation functional. Recently, large-deviation theory
has  scored sensible success in describing fluctuations in stationary
non-equilibrium regimes of various microscopic models \cite{BDGJL}.
This approach is mainly based on the extension of
the time-reversal conjugacy property introduced by Onsager and
Machlup \cite{OM} to stationary non-equilibrium states.
In practice, thermal fluctuations in irreversible stationary processes
can be traced back to a proper hydrodynamic description derived from
the microscopic evolution rules.
The general form of the action functional is
\be
I_{[(t_1,t_2)]}(\rho) = \frac{1}{2} \int_{t_1}^{t_2} dt\,\,
\langle W, K(\rho) \,\, W\rangle
\label{onsag}
\ee
where $\rho(t,\vec x)$ represents in general a vector of thermodynamic
variables depending on time $t$ and space variables $\vec x$.
The symbol $\langle \cdot \, ,\, \cdot \rangle$ denotes the integration
over space variables.
$W$ is a hydrodynamic evolution operator acting on
$\rho$: it vanishes when $\rho$ is equal to the stationary solution
$\bar\rho$, which is assumed to be unique. The positive kernel
$K(\rho)$ represents the stochasticity of the system
at macroscopic level.
According to the large deviation-theory, the entropy $S$ of a stationary
non-equilibrium state is related to the action functional $I$ as
follows:
\be
S(\rho) = \inf_{\rho} I_{[-\infty,0]} (\hat\rho)
\ee
where the minimum is taken over all trajectories connecting $\bar\rho$
to $\hat{\rho}$.

For our purposes it is enough to consider that the action functional
$I$ provides a natural measure for statistical fluctuations in
non-equilibrium stationary states, so that, formally, any statistical
inference can be obtained from $I$.
Indeed, from the very beginning we have to deal with a hydrodynamic
formulation, namely the rsNSE: in the next Section
we will argue that an action functional of the form (\ref{onsag}) can be
obtained by field-theoretic analytic calculations.

In particular,
explicit integration over all longitudinal components of the velocity
field and over the associated auxiliary fields can be performed.
This allows to obtain a hydrodynamic evolution operator $W$ which depends
only on the transverse components of the velocity field $v^{\alpha}_T(t,\vec
x)$ $(\alpha = 1, 2, 3)$. Moreover,
the positive kernel $K$ amounts to the inverse correlation function
of the stochastic source. This formulation allows to overcome some of 
the technical difficulties characterizing standard perturbative methods
and diagramatic expansions.

On the other hand, we have to face with new difficulties. 
The hydrodynamic operator appearing in the
large deviation functional is nonlinear, so that
functional integration is unfeasible. One has to
identify a solution $ {\bar v}^{\alpha}_T(t,\vec x)$ 
of the associated hydrodynamic
equation and linearize the hydrodynamic operator around
such a solution. Then, functional integration can be performed
explicitly on the ``fluctuation'' field.
In order to be well defined, this
approximate procedure would demand the uniqueness of the 
solution of the nonlinear hydrodynamic equation. 
For this reason we have restricted our choice to a class of space--time
functions which are also solutions of the linear problem.
Among them, there is only one function which satisfies physically relevant
boundary conditions (see Section III).
Statistical fluctuations have been estimated with respect
to this solution, which has also the advantage of reducing the
dependence of the generating functional on the pressure field to a 
trivial constraint.
In practice, we construct a perturbative saddle-point approach based on a
linearization procedure of the velocity field
$v^{\alpha}_T(t,\vec{x})$ around $\bar{v}^{\alpha}_T(t,\vec{x})$.
As a consequence of the nonlinear character of the original problem.
the fluctuation field 
$u^{\alpha}_T(t,\vec{x})=v^{\alpha}_T(t,\vec{x}) -
\bar{v}^{\alpha}_T(t,\vec{x})$ is found to obey a linearized hydrodynamic
problem with coefficients depending on space and time through 
$\bar{v}^{\alpha}_T(t,\vec{x})$.
It is worth stressing that even the solution of the linearized problem is 
nontrivial and it
is found to depend naturally on a perturbative parameter ${\cal R}^{-1}$,
the inverse of the Reynolds number. We exploit this property by
constructing a further perturbation procedure to obtain an explicit
expression for 
$u^{\alpha}_T(t,\vec{x})$ at different orders in ${\cal R}^{-1}$. These
points are discussed in Section IV.

Since our main purpose here is the estimation of the structure function
(see Section V)
as an average over the non-equilibrium measure induced by the action
$I$, we have to assume that the perturbative expansion applies in
a wide range of values of ${\cal R}$. In particular, we guess that it holds
also for moderately large ${\cal R}$, since a statistical average of any
observable cannot be valid for too large values of ${\cal R}$, i.e. in a
regime of fully developed turbulence. We will argue that
statistical estimates can be consistently obtained for values of ${\cal R}$
which extend up to the region of stability of the solution
$ {\bar v}^{\alpha}_T(t,\vec x) $. Beyond this region we have no practical
way of controlling the convergence of the linearization procedure.
It is worth stressing that we obtain an analytic expression of
the structure function: the so--called K41 scaling law \cite{K41} 
is recovered on a spatial
scale, whose nontrivial dependence on ${\cal R}$ is explicitly
indicated.

At the present stage, we are not able to say at which extent our 
 results on the dimensional
scaling are  dependent on the particular choice we  did for the solution 
around which  we
studied the fluctuations. Further investigations are needed to clarify this 
important point,
which probably require the combination of analytical and numerical techniques.

\section{The model}
We consider the Navier-Stokes equation for 
the velocity vector-field components $v^{\alpha}(t,\vec{x})$
describing a divergence-free homogeneous isotropic flow:
\ba
\label{NS1}
&&\left({\partial\over
\partial t} - \nu\nabla^2\right)v^{\alpha}(t,\vec{x}) +
v^{\beta}(t,\vec{x}){\partial\over \partial
x^{\beta}}v^{\alpha}(t,\vec{x}) + {1\over
\rho}{\partial\over \partial x^{\alpha}}P(t,\vec{x})  -
f^{\alpha}(t,\vec{x}) = 0,\\
&&\label{cons1}
{\partial\over \partial x^{\alpha}}v^{\alpha}(t,\vec{x}) = 0.
\ea
Here, $P$ is the pressure and the field $f^{\alpha}$ represents a 
source/sink of momentum necessary
to maintain velocity fluctuations. Customarily \cite{Niko98}, we assume
$f^{\alpha}$ to be a white-in-time zero-mean Gaussian random force with
covariance
\be
\label{ff}
\langle f^{\alpha}(t,\vec{x})f^{\beta}(t^{\prime},
\vec{x}^{\prime})\rangle = F^{\alpha\beta}\left(\vec{x} -
\vec{x}^{\prime}\right)\delta\left(t - t^{\prime}\right)\ .
\ee
Due to constraint  
(\ref{cons1}), the
field
$v^{\alpha}(t,\vec{x})$ depends only on the  transverse degrees of
freedom of
$f^{\alpha}(t,\vec{x})$. Without prejudice
of generality we can also assume divergence-free forcing, yielding the
additional relation
\be
{\partial\over \partial x^{\alpha}}F^{\alpha\beta}\left(\vec{x} -
\vec{x}^{\prime}\right)
= {\partial\over \partial x^{\beta}}F^{\alpha\beta}\left(\vec{x} -
\vec{x}^{\prime}\right) = 0\ .
\ee
A standard choice for $F^{\alpha\beta}$ is
\be
F^{\alpha\beta}(\vec{x}) ={D_0L^3\over (2\pi)^3}\int d^3p\
e^{i\vec{p}\cdot\vec{x}}(Lp)^s e^{-(Lp)^2}
{\cal P}^{\alpha\beta}(p) ,
\label{misura}
\ee
where $D_0$ is the power dissipated by the unitary mass, $p =
\left|\vec{p}\right|$,
$L$ is the integral scale, $s$ is an integer exponent (typically, $s=2$)
and 
$$
{\cal P}^{\alpha\beta}(p) = \delta^{\alpha\beta}-{p^{\alpha}p^{\beta}\over 
p^2}
$$  
is the projector on the transverse degrees of freedom.

Following the Martin-Siggia-Rose formalism \cite{MSR} we introduce
the Navier-Stokes density of Lagrangian
\ba
{\cal L}(v, w, P, Q, f) &&= w^{\alpha}(t,\vec{x})\left[
\left({\partial\over
\partial t} - \nu\nabla^2\right)v^{\alpha}(t,\vec{x}) +
v^{\beta}(t,\vec{x}){\partial\over \partial
x^{\beta}}v^{\alpha}(t,\vec{x})\right.\nonumber\\ 
&&\left.+ {1\over
\rho}{\partial\over \partial x^{\alpha}}P(t,\vec{x}) -
f^{\alpha}(t,\vec{x})\right] + {1\over
\rho}Q(t,\vec{x}) {\partial\over \partial
x^{\alpha}}v^{\alpha}(t,\vec{x})\ ,
\ea
where the field $w^{\alpha}$ is the conjugated variable to the velocity field
$v^{\alpha}$ and the field $Q$ is the Lagrangian multiplier related to
constraint (\ref{cons1}). The generating functional is given by
the integral
\ba
{\cal W}\left(J, P\right)   
&&= \int {\cal D}v{\cal D}w{\cal D}Q{\cal D}f
\exp\left\{i\int dt\ d^3x\left[{\cal L}(v, w, P, Q, f) +
J_{\alpha}v^{\alpha}\right]\right. \nonumber\\
&&\left.-{1\over 2}\int dt
d^3xd^3yf^{\alpha}F^{-1}_{\alpha\beta}f^{\beta}\right\}
\label{func01}
\ea
where $J_{\alpha}$ are the components of the "external source" vector $J$.
By integration over the statistical measure, ${\cal D}f e^{-{1\over 2}\int 
fF^{-1}f}$ and over the Lagrange multiplier $Q$, we obtain an expression 
which depends only on the transverse component $v_T$ of
the velocity field $v$. By decomposing the auxiliary field $w$ in terms of 
its transverse ($w_T$) and longitudinal ($w_L$) components, $w =
w_L + w_T$, the measure ${\cal D}w$ factorizes into 
${\cal D}w_L{\cal D}w_T$
and the generating functional (\ref{func01}) reduces to
\ba
{\cal{W}}(J, P) = \int{\cal{D}}w_T{\cal{D}}w_L{\cal{D}}v_T\ exp\left\{
i\int dtd^3x\left[w^{\alpha}_T\left\{\left({\partial\over \partial t}
-\nu\nabla^2\right) v_{\alpha T} + \left(v^{\beta}_T{\partial\over \partial
x^{\beta}}v_{\alpha T}\right)_T\right\}\right.\right.\nonumber\\
\left.\left.+ w^{\alpha}_L\left\{\left(v^{\beta}_T{\partial\over \partial
x^{\beta}}v_{\alpha T}\right)_L + {1\over \rho}{\partial P\over \partial
x^{\alpha}}\right\} + J_{\alpha}v^{\alpha}_T\right]-{1\over 2}\int dt\int 
d^3xd^3y w^{\alpha}_TF_{\alpha\beta}w^{\beta}_T\right\}\ .
\label{funct00}
\ea
Diagramatic strategies are usually applied at this level. We want to point
out that one can go further by observing
that also the transverse and longitudinal components of the auxiliary field
$w$ can be integrated out, yielding the equation
\be
{\cal W}(J, P) = \int {\cal D}v_T e^{-{1\over 2}I(v_T) + i\int dt d^3x
J_{\alpha}v^{\alpha}_T}
\delta\left(\left(v^{\beta}_T{\partial\over \partial
x^{\beta}}v_{\alpha T}\right)_L + {1\over \rho}{\partial P\over \partial
x^{\alpha}}\right)
\label{funct1}
\ee 
where the action functional $I$ has the form
\ba
I(v_T) &&= \int dt d^3x d^3y\left[\left({\partial\over \partial t} -
\nu\nabla^2\right)v_T^{\alpha}(t, \vec{x}) + v_T^{\rho}(t, \vec{x})
\partial_{\rho}v_T^{\alpha}(t, \vec{x})\right]\nonumber\\
&&F^{-1}_{\alpha\beta}(|\vec{x}-\vec{y}|)
\left[\left({\partial\over \partial t} -
\nu\nabla^2\right)v_T^{\beta}(t, \vec{y}) + v_T^{\lambda}(t, \vec{y})
\partial_{\lambda}v_T^{ \beta}(t, \vec{y})\right]\ .
\label{act}
\ea
The computation of (\ref{funct1}) would require to solve the constraint
\be
\left(v^{\beta}_T{\partial\over \partial
x^{\beta}}v_{\alpha T}\right)_L + {1\over \rho}{\partial P\over \partial
x^{\alpha}} =0\ ,
\label{press1}
\ee
In principle, this is a very difficult task due to the nonlinear character
ot the constraint.\\
In the following section we show that we can identify a particular 
extremal solution, $\bar{v}_T$, of the functional (\ref{act}). This solution 
is found
to be independent of the stochastic source and, moreover, it satisfies
constraint (\ref{press1}) for any constant value of the pressure.
Accordingly,  $I(v_T)$ can be interpreted as a large deviation functional
(see eq.(1))
and the statistical nonequilibrium measure of the rsNSE can be 
effectively evaluated by integrating over the fluctuations around this
extremal solution.
It is worth observing that
the entropy is related to the functional
$I(v_T)$ by the relation \cite{BDGJL}
\be
S(v_T) = \frac{1}{2}\inf_{v_T}I(v_T )\ ,
\ee
where the minimum is taken over all trajectories connecting  
$\bar{v}_T$ to $v_T$ .\\ In what follows we are
going to show that a suitable perturbative strategy can be applied for
obtaining explicit analytic calculations of the statistical properties 
of the rsNSE.

\section{A quasi-steady solution and its stability}
Any analytic approach aiming at the estimation of the generating functional
(\ref{funct1}) demands the identification of an explicit solution of
the action functional (\ref{act}). In practice, this amounts to solve
the stationarity condition
\ba
{\delta I(v_T)\over  \delta v^{\sigma}_T(t, \vec{x})} &&= 2\int d^3y\left[
-\delta_{\sigma}^{\ \alpha}\left({\partial\over \partial t} +
\nu\nabla^2\right) + \partial_{\sigma}v_T^{\alpha}(t,
\vec{x})\right.\nonumber\\
&&-\delta_{\sigma}^{\ \alpha}v_T^{\rho}(t,
\vec{x})\partial_{\rho}\bigg]\left(F^{-1}_{\alpha\beta}(|\vec{x}-\vec{y}|)
\left[\left({\partial\over \partial t} -
\nu\nabla^2\right)v_T^{\beta}(t, \vec{y})\right.\right.\nonumber\\
&&+ v_T^{\lambda}(t, \vec{y})
\partial_{\lambda}v_T^{ \beta}(t, \vec{y})\bigg]\bigg) = 0\ .
\label{estremo}
\ea
We want to observe that for any arbitrary scalar field
$\Phi(t, \vec{x})$ a solution of the equation
\be
\left({\partial\over \partial t} -
\nu\nabla^2\right)v_T^{\beta}(t, \vec{x}) + v_T^{\lambda}(t, \vec{x})
\partial_{\lambda}v_T^{ \beta}(t, \vec{x}) =
\partial^{\beta}\Phi(t, \vec{x})\ ,
\label{ridotta}
\ee
is also a  solution of (\ref{estremo}). 
Since $F^{\alpha\beta}(|\vec{x}-\vec{y}|)$ contains a projector on the
transverse degrees of freedom we can fix, 
without prejudice of generality,
the condition $\partial^{\beta}\Phi=0$.
It is worth pointing out that, for what concerns eq.(\ref{ridotta}), 
this condition implies also that the longitudinal component of the 
nonlinear term vanishes, i.e. the solution 
$v_T^{\lambda}(t, \vec{x})$ has to satisfy the additional condition
\be
\partial_{\beta}\left(v_T^{\lambda}(t, \vec{x})
\partial_{\lambda}v_T^{ \beta}(t, \vec{x})\right) = 0\ .
\label{cond2} 
\ee
Several different solutions can be found: among them, the only one
unaffected by divergences in space and time is the following:
\be
\bar{v}^{\alpha}_T(t, \vec{x}) = {U^{\alpha}\over 2}\left\{1 + e^{-{t\over
\tau_D}}\sin\left({1\over
2\sqrt{b^2-(\vec{a}\cdot\vec{b})^2}}
\left(\vec{b}\wedge\vec{a}\right)\cdot{\vec{x}\over L}\right)\right\}\ ,
with \quad t>0\ .\\
\label{soluz}
\ee
The $U^{\alpha}$ are the components of the vector of velocity amplitude 
$\vec{U}$ ($U=|\vec{U}|$)~,
$\vec{a} = {\vec{U}\over U}$ is the corresponding 
unit vector and $\vec{b}$ identifies a rotation axis.  
Both vectors $\vec{U}$ and $\vec{b}$ can be fixed arbitrarily.
We assume also that the length--scale $L$ is the same as the forcing 
integral scale defined in (\ref{misura}). This implies that 
solution (\ref{soluz}) decays exponentially in time to the constant
${U^{\alpha}\over 2}$ with the rate $\tau_D=4L^2/\nu$, which is the 
diffusion time scale. The dependence of solution (\ref{soluz}) 
on the Reynolds number ${\cal R}$ can be made explicit by the
relation ${\cal R} = {LU\over \nu}$, so that 
$\tau_D= 4\nu{\cal R}^2/U^2$.  
Notice that condition
(\ref{cond2}) is trivially satisfied by solution (\ref{soluz}), because
\be
\bar{v}^{\beta}_T(t, \vec{x})\partial_{\beta}\bar{v}^{\alpha}_
T(t, \vec{x}) = 0\ .
\label{triv}
\ee
Accordinlgy, $\bar{v}^{\alpha}_T(t, \vec{x})$ is also a solution of the 
diffusion equation
$\left(\partial_t-\nu\nabla^2\right)\bar{v}_T^{\alpha}(t, \vec{x}) = 0$.
There are two main consequences to be pointed out: i) as a 
solution of the linear diffusion equation $\bar{v}^{\alpha}_T(t, \vec{x})$
is unique, which is a crucial requirement for the large
deviation approach; ii) the solution has to be defined only for positive 
times.\\
Moreover, due to condition (\ref{triv}), the constraint (\ref{press1}) 
is trivially solved by $P=constant$.\\

For $|\vec{x}| \ll L$ solution (\ref{soluz}) approximates a linear shear 
flow: this is well known to produce instabilities for sufficiently 
large Reynolds numbers ${\cal R}$. In this perspective, it is worth analyzing
the dynamical stability of (\ref{soluz}). For this aim we consider the
perturbed velocity vector, whose components are:
\be
v^{\alpha}(t, \vec x) =
\bar{v}^{\alpha}_T(t, \vec x) + \delta v^{\alpha}_T(t, \vec x)
\label{perturb}
\ee
The perturbation vector $\delta {v}^{\alpha}_T$ is
assumed to be much smaller than $\bar{v}^{\alpha}_T$ with respect to
any proper functional measure $\mu$ , i.e. $ 
|\delta v^{\alpha}_T(t, \vec x)|_{\mu} << 
|v^{\alpha}(t, \vec x)|_{\mu} \,\,\, , \forall t$ and 
$\forall \vec x$~. One can substitute (\ref{perturb}) into
(\ref{ridotta}) with $\partial^{\beta}\Phi=0$, while
assuming that it satisfies constraint (\ref{cond2}).
In the linear approximation one obtains an equation for 
$\delta v^{\alpha}_T(t, \vec x)$, which can be solved explicitly
by performing an expansion in the inverse Reynolds number $
{\cal R}^{-1}$. As shown in Appendix A, 
the perturbation field vanishes and, accordingly, (\ref{soluz}) is stable
for sufficiently large times and 
Reynolds numbers and provided the following inequality
holds:
\be
{8\nu^2{\cal R}\over U^2}k^2> 1\ .
\label{cond1}
\ee

This inequality implies that for increasing values of $\cal R$ the band 
of unstable modes becomes thinner and thinner. As a consequence, solving the 
stability problem by expanding the solution of the linearized dynamics 
(\ref{2-5}) in powers of ${\cal R}^{-1}$ is consistent with this
finding. Since condition (\ref{cond1}) has been derived by assuming 
${\cal R}$ large, it is not in contradiction with the Landau 
scenario for the origin of turbulence.

In summary, $\bar{v}^{\alpha}_T(t, \vec{x})$ exhibits all
the expected features of a physically relevant solution, which
corresponds to stationarity conditions for the large--deviation
functional. Accordingly, it can be effectively used 
for computing statistical non--equilibrium fluctuations of the
rsNSE. In the next section we will exploit
a saddle point strategy for performing explicit calculations
from the generating functional.

\section{Perturbative analysis of the generating functional}
All statistical properties concerning the rsNSE are contained in
the structure functions which can be obtained by performing
derivatives of the generating functional (\ref{funct1}) with respect to
the current $J^{\alpha}$. 
An explicit calculation is
unfeasible due to the nonlinear character of the action functional 
$I(v_T)$. Since in the previous section we have identified the 
solution $\bar{v}^{\alpha}_T$, we can tackle the problem  
by introducing the velocity field $u^{\alpha}_T = v^{\alpha}_T - 
\bar{v}^{\alpha}_T $,
which represents fluctuations with respect to $\bar{v}^{\alpha}_T$, and
by applying a saddle--point strategy.\\

Due to the translational invariance of the functional measure,
the generating functional (\ref{funct1}) can be rewritten as
\be
{\cal W}(J) = \int{\cal D}u_T\ e^{-{1\over 2}I(u_T) +i\int
dtd^3x J_{\alpha} u^{\alpha}_T} .
\label{AF1}
\ee
A linearized expression for the action functional can be obtained by
assuming that higher order terms in $u_T^{\alpha}$
generated by the saddle-point expansion around the solution
$\bar{v}^{\alpha}_T$ are negligible with respect
to the functional measure ${\cal D}u_T$:
\ba
I(u_T) &&= \int_0^{\infty} dt \int 
d^3xd^3y\left[(\partial_t-\nu\nabla^2_x)u^{\alpha}_T(\hat{x})
+\bar{v}^{\rho}_T
(\hat{x}) \partial_{\rho}u^{\alpha}_T(\hat{x})
+ u^{\rho}_T(\hat{x})\partial_{\rho}
\bar{v}^{\alpha}_T
(\hat{x})\right]
F^{-1}_{\alpha\beta}(|\vec{x}-\vec{y}|)\times\nonumber\\
&&\left[(\partial_t-\nu\nabla^2_y)u^{\beta}_T(\hat{y})
+\bar{v}^{\lambda}_T
(\hat{y})\partial_{\lambda}u^{\beta}_T(\hat{y})
+ u^{\lambda}_T(\hat{y})\partial_{\lambda}
\bar{v}^{\beta}_T
(\hat{y})\right] 
\ .
\label{FF}
\ea
We have also introduced the shorthand notation
$\hat{x}\equiv (t, \vec{x})$.\\
Consistently with this
perturbative approach, we can also assume that, at leading order,
constraint (\ref{press1}) is still trivially  solved by (\ref{soluz}), i.e.
the pressure $P$ is a constant.

In this way the action functional (\ref{FF}) has a 
bilinear form in the field $u_T^{\alpha}$, with coefficients depending
on $\bar{v}^{\alpha}_T$. In order to perform explicit
Gaussian integration of the generating functional one has first to
understand how the technical difficulties inherent such a dependence
can be circumvented. The first problem that we have to face with
is that, since (\ref{soluz}) is defined only for $t>0$, also 
(\ref{FF}) is defined for positive times. As we discuss in Appendix B,
a standard procedure allows one to get rid of any singularity of the action
integral that might emerge for $t \to 0^+$. This is a consequence of the
stucture of the linearized hydrodynamic operator appearing in
(\ref{FF}). The second problem concerns the possibility of obtaining an
analytic expression for the generating functional. To this aim one can
exploit a perturbative expansion of (\ref{soluz}) in powers of the inverse
Reynolds  number
${\cal R}^{-1}$.  Actually, it is worth rewriting the solution 
(\ref{soluz})  making explicit its dependence on the
Reynolds number: 
\be
\bar{v}^{\alpha}_T(t, \vec{x}) = {U^{\alpha}\over 2}\left\{1 + e^{-{U^{2}\over
4\nu{\cal R}^2}t}\sin\left({2\over
\sqrt{b^2-(\vec{a}\cdot\vec{b})^2}}
{\left(\vec{b}\wedge\vec{U}\right)\cdot\vec{x}\over 4\nu{\cal
R}}\right)\right\}\ ,
with \quad t>0\ .\\
\label{soluz1}
\ee
Using ${\cal R}^{-1}$ as a perturbative parameter, one can expand
$\bar{v}^{\alpha}_T$ at all orders in 
${\cal R}^{-1}$. When this expansion is substituted 
into (\ref{FF}) at leading order 
the action functional, in Fourier transformed variables, takes the form
\be
I(u_T) = \int {d^4p\over (2\pi)^4}u^{\rho}_T(-\hat{p})M_{\rho}^{\ 
\alpha}(-\hat{p})
F^{-1}_{\alpha\beta}(p)
M^{\beta}_{\ \zeta}(\hat{p})
u^{\zeta}_T(\hat{p}) + O\left({1\over {\cal R}^2}\right)\ .
\label{act1}
\ee
We denote with $u^{\zeta}_T(\hat{p})$ the Fourier
transform of the field $u^{\zeta}_T(\hat{x})$ with
$\hat{p}\equiv (p_0,\vec{p})$, $p_0$ and $\vec{p}$ being
the Fourier--conjugated variables of $t$ and $\vec{x}$, respectively.
We introduce the representation of the action functional in terms of
the Fourier--transformed variables because this makes more transparent 
the diagonalization procedure required to arrive at the final result.
\\
The hydrodynamic evolution term
$M^{\beta}_{\ \zeta}(\hat{p})u^{\zeta}_T(\hat{p})$ is given
by the expression
\be
M^{\beta}_{\ \zeta}(\hat{p})u^{\zeta}_T(\hat{p})
= \left\{\delta^{\beta}_{\ \zeta}\left[i\left(p_0 + {1\over 
2}\vec{p}\cdot \vec{U}\right) +\nu
p^2 -{C\over 4}\vec{p}\cdot\vec{U} 
{\left(\vec{b}\wedge\vec{U}\right)^{\gamma}\over 4\nu{\cal
R}}\partial_{p_{\gamma}}\right]  -{C\over
4}U^{\beta}{\left(\vec{b}\wedge\vec{U}\right)_{\zeta}\over
4\nu{\cal R}}\right\}u^{\zeta}_T(\hat{p}) ,
\label{matr1}
\ee
where $C = {2\over \sqrt{b^2-(\vec{a}\cdot\vec{b})^2}}$.
\\
The next step in this calculation requires the diagonalization of the
matrix 
$M_{\rho}^{\ \alpha}(-\hat{p}){F^{-1\, \alpha\beta}(p)}M^{\beta}_{\
\zeta}(\hat{p})$. 
Since by definition the factor $[F^{\alpha\beta}(p)]^{-1}$ is 
proportional to the identity operator in the space of the transverse 
solutions\footnote{More explicitly we have $[F^{\alpha\beta}(p)]^{-1} =
F^{-1}(p)
{\cal P}^{\alpha\beta}(p)$ where $F(p) = D_0L^3(Lp)^s e^{-(Lp)^2}$} we have 
just
to diagonalize the matrix of the hydrodynamic operator
$M^{\beta}_{\ \zeta}(\hat{p})$.\\
The computation  of the eigenvalues, $\lambda_1$, $\lambda_2$ and
$\lambda_3$ of
$M^{\beta}_{\ \zeta}(\hat{p})$ deserves lengthy calculations sketched in
Appendix \ref{C}.   Hereafter,  we report the final form of the generating
functional:
\be
{\cal W}(\eta) = \int {\cal J}(H){\cal D}\phi_T\ e^{-{1\over 
2}\int_{\hat{p}}\phi_T^{\rho}(-\hat{p})F^{-1}(p)
I_{\rho\gamma}(\hat{p})\phi_T^{\gamma}(\hat{p})
+i\int_{\hat{p}}
\eta_{T\alpha}(-\hat{p})\phi_T^{\alpha}(\hat{p})}\ ,
\ee
where we have used the shorthand notation $\int_{\hat{p}}\equiv \int
{d^4p\over (2\pi)^4}$ and
\be
I_{\rho\gamma}(p) = \left(\begin{array}{ccc}
\lambda_1^*(\hat{p})\lambda_1(\hat{p}) & 0 & 0
\\
0 & \lambda_2^*(\hat{p})\lambda_2(\hat{p}) & 0
\\
0 & 0 &\lambda_3^*(\hat{p})\lambda_3(\hat{p})
\end{array}
\right)\ ,
\ee
${\cal J}(H)$  is the Jacobian of the basis transformation
$u\longrightarrow \phi$, $J\longrightarrow \eta$ engendered by
the matrix $H$, which diagonalizes
$M_{\rho}^{\ \alpha}(\hat{p})$.
It is worth pointing out that
the transformed vector $\phi_T^{\alpha}(\hat{p})$ still represents
transverse components.
Gaussian integration yields the following expression of the
{\it normalized} functional in terms of the $\eta^{\alpha}$ source fields
\be
{\cal W}(\eta) = e^{-{1\over 
2}\int_{\hat{p}}\eta^{\rho}_T(-\hat{p})F(p)I^{-1}_{\rho\gamma}
\eta^{\gamma}_T(\hat{p})}\ .
\label{FJ1}
\ee
In practice, the explicit computation of the structure functions can be
accomplished by returning to the original representation, where the
generating functional has the form
\be
{\cal W}(J) = e^{-{1\over 2}\int_{\hat{p}}J^{\rho}_T(-\hat{p})F(p)
\left(HI^{-1}H^T\right)_{\rho\sigma}(\hat{p})J^{\sigma}_T(\hat{p})}.
\label{FJ2} 
\ee
In the next section we are going to derive an explicit expression for the 
second--order structure function.

\section{Short-distance behavior of the
second order structure function}
The analytic expression obtained for 
the generating functional (\ref{FJ2}) allows one to obtain
all the statistical information about the  fluctuations
around the basic solution $\bar{v}^{\alpha}_T$. In this section
we perform the explicit calculation of the 
second-order structure function of the
velocity field $u^{\alpha}$, defined as
\ba
S_2&&=\langle\left|u_T(t, \vec{r}+\vec{x})-u_T(t,\vec{x})\right|^2
\rangle\nonumber\\
&&= \langle\left|(u^{\alpha}_T(t,
\vec{r}+\vec{x})-u^{\alpha}_T(t, \vec{x}))(u_{T\alpha}(t,
\vec{r}+\vec{x})-u_{T\alpha}(t, \vec{x}))\right|\rangle ,
\label{struttura}
\ea
The brackets denote averages over the stochastic forcing.\\
By assuming isotropy and homogeneity of the velocity field $u^{\alpha}$,
expression (\ref{struttura}) is expected to assume the typical
form of a scale invariant function
\begin{equation}
S_2(r) =  r^{\zeta_2}F_2\left(t, {r\over L}\right) 
\end{equation}
Here $r = |\vec{r}|$ and $L$ is the integral scale associated with the
noise source. It is worth stressing that, at variance with
fully developed turbulent regimes, here the assumption of
isotropy and homogeneity have to be taken as a plausible
hypothesis allowing for analytic computations.\\

We want to point out that any exponent $\zeta_n$ must be
independent of the basis chosen for representing
the functional ${\cal W}$. For the sake of simplicity, it is worth using
(\ref{FJ1}) rather than (\ref{FJ2}) to obtain:
\be 
S_2(r) =\left.\left({\delta\over
i\delta\eta^{\alpha}_T(t,\vec{x}+\vec{r})} - {\delta\over
i\delta\eta^{\alpha}_T(t,\vec{x})}\right)
\left({\delta\over i\delta\eta_{T\alpha}(t,\vec{x}+\vec{r})}
- {\delta\over i\delta\eta_{T\alpha}(t,\vec{x})}\right){\cal
W}(\eta)\right|_{\eta=0}\ .
\label{S21}
\ee
As shown in Appendix \ref{D}, it turns out that $S_2(r)$ can be
rewritten as follows:
\begin{equation}
{S}_2(r) =-{1\over \nu}\left(I_1(r) +I_2(r)\right) .
\end{equation}
where 
\be
I_1(r) = {D_0\over 
(2\pi)^2}r^2\sum_{n=0}^{\infty}(-1)^{n+1}{\Gamma\left({s+3+2n\over
2}\right)\over
\Gamma\left(2n+4\right)}\left({r\over L}\right)^{2n}\ 
\label{i11}
\ee
and
\ba
I_2(r) &&= D_0L^3{32\nu^2{\cal R}\over 
U^2}\left\{\int_0^{\infty}{p^2dp\over 
(2\pi)^2}(Lp)^se^{-(Lp)^2}\int_{-1}^1dx\left(e^{iprx} -
1\right)\right.\nonumber\\
&&\left.\times\left(\sum_{l=1,2}{\left(1-x^2\right)^{1\over 3}\over 
x^{2\over 3}}
{\left[\sum_{m=0}^2s_{lm}
F_m\left(x, {8\nu^2{\cal R}\over U^2}p^2; \Sigma, \Xi\right)
 + {1\over 2}\Sigma{x^{2\over 3}\over \left(1-x^2\right)^{1\over 
3}}\right]\over
\prod_{i\not= l}\left(\sum_{k=0}^2\left(s_{lk}
-s_{ik}\right)F_k\left(x, {8\nu^2{\cal R}\over U^2}p^2; \Sigma, 
\Xi\right)\right)}
+ O\left({1\over {\cal R}^2}\right)\right)\right\} . \nonumber\\
\label{tremendo}
\ea
The coefficients $s_{ij}$ and the functions $F_i$, together with
their arguments, are specified in Appendix \ref{D} .\\
The main contribution of the stochastic measure $p^{2+s}e^{-(L p)^2}dp$ 
to the first integral in (\ref{tremendo}) comes from a narrow region of
wavenumbers close to $\bar{p}$, where the function $p^{2+s}e^{-(Lp)^2}$ has
its maximum, i.e.
\be
\bar{p}={1\over L}\sqrt{s+2\over 2}\ .
\ee
Accordingly, the function ${8\nu^2{\cal R}\over U^2}p^2$ contributes to the
integral by taking values close to $4(s+2)\over {\cal R}$.\\
Moreover, for $p=\bar{p}$ the sufficient condition (\ref{cond1}) for the 
stability of small perturbations determines an upper bound for the
Reynolds number:
\be
{\cal R}
\lesssim 4(s+2)\ ,
\label{stabi}
\ee
This implies that for sufficiently small $ {\cal R} $ the 
wavenumber $\bar{p}$ is stable. Under this condition, the leading 
contribution in (\ref{tremendo}), consistently with the expansion in 
${\cal R}^{-1}$, can be obtained
by performing an expansion in powers of $U^2\over 8\nu^2{\cal R}p^2$.\\
One finally obtains the complete expression of the structure function
(see Appendix {D} for details)
\ba
{S}_2(r) &&= -{1\over \nu}\left(I_1(r)+I_2(r)\right)\nonumber\\
&&\sim -{D_0\over
(2\pi)^2\nu}r^2\sum_{n=0}^{\infty}\left\{(-1)^{n+1}\Gamma\left({s+2n+3\over 
2}\right)
\left[{1+\Xi\over \Gamma\left(2n + 4\right)}
-{2^{13\over 3}\Xi\over \Sigma^{2\over 3}}{2n+4\over
\Gamma (2n+6)}\right]\left({r\over L}\right)^{2n}\right\}\ ,\nonumber\\
&&\quad
\mbox{for} \quad 1<{\cal R}\ll 4(2+s)\ ,
\label{SS2}
\ea
At leading order in the distance $r$ this expression is dominated by a 
dissipative contribution.

We conjecture that this analysis can be extended
to the parameter region defined by the condition ${\cal R}\gtrsim 4(2+s)$,
where  the statistically relevant wavenumbers can be unstable.
As shown in Appendix \ref{D}, in this case $I_2(r)$ has two contributions: one
is again dissipative, while there is another one yielding the nontrivial
scaling behavior $r^{2/3}$.
Specifically, the expression of $S_2(r)$ for ${\cal R}\gtrsim 4(2+s)$
is found to be 
\ba
{S}_2(r)&&\sim -{D_0\over \pi\nu}\left\{{1+{\Xi\over 2}\over
4\pi}r^2\sum_{n=0}^{\infty}(-1)^{n+1}{\Gamma
\left({s+2n+3\over 2}\right)\over \Gamma(2n+4)}
\left({r\over L}\right)^{2n}\right.\nonumber\\
&&\left.+ {{\cal R}^{1\over 3}\over \Gamma\left({2\over 3}\right)}
\left({\nu\over U}\right)^{4\over 3}r^{2\over 3}
\sum_{n=0}^{\infty}C_n(\Sigma)\Gamma\left({3s+3n+5\over 
6}\right)\left({r\over
L}\right)^n  \right\}
\label{last}
\ea
This expression is dominated by the term $r^{2/3}$ for sufficiently small 
distances. 
Indeed, 
the crossover scale between the $r^2$ and the $r^{2\over 3}$ terms
occurs at
\be
{r\over L}\sim F{\cal R}^{-{3\over 4}}\ .
\ee
In Appendix \ref{D}  we evaluate the constant $F\sim 0.6$ and we report
the  expression of the numerical coefficient $C_0(\Sigma)$. The general
expression of the coefficients $C_n(\Sigma)$ appearing in (\ref{last}) 
has been omitted, because it has no practical interest for explicit
calculations.

It is a remarkable fact that $S_2$ can exhibit the scaling 
behavior predicted by the K41 theory, which is assumed to hold (apart
from intermittency corrections)
when the velocity fluctuations are turbulent
in the so-called inertial range of scales. This suggests
that hydrodynamic fluctuations in a system at the very
initial stage of instability development already contain
some properties attributed to the developed turbulence regime.

\section{Conclusions}

In this paper we have exploited the field-theoretic approach
to reformulate the random forced Navier--Stokes
problem in terms of the evaluation of a quadratic action. 
This has the formal structure of a large--deviation functional,
describing thermal fluctuations of irreversible stationary processes.
The crucial step for obtaining such a statistical representation
is the integration over all longitudinal components
of both velocity and the associated auxiliary fields. With respect
to the standard formulation which yields usual
diagramatic strategies, we perform one more field
integration.
The positive definite kernel, which connects the hydrodynamic
evolution operator in the action functional, 
is the inverse of the forcing correlation function.\\
In terms of the action functional, the knowledge of the
whole velocity statistics
reduces to the computation of functional integrals. However,
due to the intrinsic nonlinear character of the hydrodynamic
operator several technical difficulties have been solved 
for performing analytic calculations.
In particular, one has to to introduce suitable approximations.\\
In order to obtain an analytic expression of the generating functional 
we have identified a solution
around which we have linearized the hydrodynamic evolution operator. 
We have also introduced a velocity field which represents 
fluctuations with respect to this solution. A perturbative
expansion in the inverse Reynolds number finally yields the wanted
result.

In principle, from this analytic treatment one can obtain all relevant 
statistical
information about the rsNSE by computing any velocity multipoint
structure function. In this paper we report only the explicit
calculation of the two--point second order moment of the velocity
field. 
As shown in the Appendices, the algebraic manipulations needed for
obtaining the final result are far from trivial also in this simple case.

In fact, in this paper we aim at understanding whether fluctuations at the
early stage of their development (accordingly, we dub them as pre-turbulent
fluctuations) already contain some important features of developed
turbulence. We are interested, in particular, to characterize the
scale invariant properties of such fluctuations.
In this respect, we find that they are organized at
different scales in a self--similar way. Remarkably, the scaling exponent
coincides with the dimensional prediction of the Kolmogorov 1941
theory \cite{K41}, valid for developed turbulence regimes. Whether or not such
exponent is a genuine reminiscence of the developed turbulence
phenomenology needs further investigations. \\
Unfortunately, the complexity of the derivation leading to the
K41 scaling law does not
allow us to identify precisely the very origin of such a dimensional
prediction. We can however argue a relationship between the
observed dimensional scaling and the conservation laws (for momentum
and energy) associated with the two eigenvalues of the matrix appearing
in the action functional (\ref{act1}).\\
Finally, it is worth observing that the dimensional scaling law
emerges for a particular choice we did for the pressure field:
fluctautions have been restricted around a solution for which 
the pressure is constant. Unfortunately, owing to the fact that the
analytical  treatment is not duable in the general case,
we cannot substantiate the fact on whether 
the dimensional prediction we found  is not a  consequance  of our
particular choice for the pressure fields.\\
At least three scenaries might be possible. Firstly, pressure field does
not affect neither the leading (dimensional) scaling law nor its
prefactor. 
It only affects the subleading scaling contributions.
In this case our  simplification would capture  the relevant physics of
the 
problem. The second possibility is that the leading scaling law does not
change
but this is not for the prefactor. The last possibility is that pressure 
changes the (domensional)  scaling law giving rise to intermittency 
corrections. 
Unfortunately, at the present stage of our knowledge, we are not in the
position to select
one  scenary among the three we have pointed out. Further investigations
are needed for this aim, which probably call to deep numerical
investigations of the system under consideration.
\\
We want to conclude by outlining some open problems and perspectives.
A first question concerns the physical relevance of 
the solution (\ref{soluz}) around which we linerize the evolution
operator.  It represents a shear-like solution, which is a well-known
generator of instability. Moreover,
its unicity and stability properties seem to indicate that this solution
can play a major role in the determination of stationary nonequilibrium 
fluctuation statistics to be attributed to the rsNSE. As a mathematical
object, it exhibits all the wanted features that one would like to
attribute to such a solution. On the other hand, the authors have not
yet a physical intuition for its relevance and aim at making some future
progress in this direction.\\
Another interesting point to be tackled concerns the computation of the
third-order momentum of the velocity correlators. In this case the
predictions of our approach could be compared with the $4/5$-law, which
is one among the very few exact results of turbulence theories.\\
Finally, the extension of our results to other classes of transport
problems, including passive scalar advection, could provide
a better understanding of the basic mechanism at the origin of
the observed scaling behaviors.

\begin{acknowledgments}
This work has been supported by Cofin 2003 
``Sistemi Complessi e Problemi a Molti Corpi'' (AM).
We acknowledge useful discussions with G. Jona-Lasinio, M. Vergassola,
P. Constantin and P. Muratore--Ginanneschi.
\end{acknowledgments}

\begin{appendix}
\section{}
\label{A}

In this Appendix we perform the stability analysis of the solution
$\bar{v}^{\alpha}_T$  by the linearized equation
\be
\left({\partial\over \partial t} - \nu\nabla^2\right) \delta 
v^{\gamma}_T(t,\vec{x}) +
\bar{v}_T^{\beta}(t,\vec{x}){\partial\over \partial x^{\beta}}
\delta v^{\gamma}_T(t,\vec{x}) + \delta v^{\beta}_T(t,\vec{x}){\partial\over 
\partial
x^{\beta}} \bar{v}_T^{\gamma}(t,\vec{x}) = 0\ ,
\label{2-5}
\ee
with the constraint
$$
{\partial\over \partial 
x^{\gamma}}\left(\bar{v}_T^{\beta}(t,\vec{x}){\partial\over \partial 
x^{\beta}}
\delta v^{\gamma}_T(t,\vec{x}) + \delta va^{\beta}_T(t,\vec{x}){\partial\over 
\partial
x^{\beta}} \bar{v}_T^{\gamma}(t,\vec{x})\right) = 0\ .
$$ 
In Section III we have already observed that
$\bar{v}^{\alpha}_T$ is a quasi-steady solution for a time \
$t\ll\tau_D={4\nu{\cal R}^2\over
U^2}$.
The Fourier transform of eq.(\ref{2-5})
with respect to the space vector $\vec{x}$ yields:
\ba
&&{\partial\over \partial t}\delta \tilde{v}^{\alpha}_T\left(t,
\vec{k}\right) - \nu k^2\delta \tilde{v}^{\alpha}_T\left(t, \vec{k}\right)
+{i\over 2}\vec{k}\cdot\vec{U}\delta \tilde{v}^{\alpha}_T\left(t,
\vec{k}\right) + {1\over 4}e^{-{t\over \tau_D}}\left\{U^{\beta}k_{\beta}
\left[\delta \tilde{v}^{\alpha}_T\left(t, \vec{k}
- C{\vec{b}\wedge\vec{U}\over 4\nu{\cal R}}\right)\right.\right.\nonumber\\
&&\left.\left.- \delta \tilde{v}^{\alpha}_T\left(t, \vec{k}
+ C{\vec{b}\wedge\vec{U}\over 4\nu{\cal R}}\right)\right]
+ U^{\alpha}C{\left(\vec{b}\wedge\vec{U}\right)_{\beta}\over 4\nu{\cal R}}
\left[\delta \tilde{v}^{\beta}_T\left(t, \vec{k} -
C{\vec{b}\wedge\vec{U}\over  4\nu{\cal R}}\right)
+ \delta \tilde{v}^{\beta}_T\left(t, \vec{k}
+ C{\vec{b}\wedge\vec{U}\over 4\nu{\cal R}}\right)\right]\right\}\nonumber\\
&&= 0\ ,
\ea
where $C = {2\over \sqrt{b^2-(\vec{a}\cdot\vec{b})^2}}$.
By performing a perturbative expansion up to second order in the parameter 
${\cal R}^{-1}$, one obtains the system of equations
\ba
\label{AA1}
&&{\partial\over \partial t}\delta \tilde{v}^{\alpha}_{T(0)}\left(t, 
\vec{k}\right) + \nu k^2\delta \tilde{v}^{\alpha}_{T(0)}\left(t,
\vec{k}\right) +{i\over 
2}\vec{k}\cdot\vec{U}\delta \tilde{v}^{\alpha}_{T(0)}\left(t, \vec{k}\right)
=  0\ ,\\
\label{AA2}
&&{\partial\over \partial t}\delta \tilde{v}^{\alpha}_{T(1)}\left(t, 
\vec{k}\right) + \nu k^2\delta \tilde{v}^{\alpha}_{T(1)}\left(t,
\vec{k}\right) +{i\over 
2}\vec{k}\cdot\vec{U}\delta \tilde{v}^{\alpha}_{T(1)}\left(t, 
\vec{k}\right)\nonumber\\
&&={1\over 
2}\vec{k}\cdot\vec{U}C{\left(\vec{b}\wedge\vec{U}\right)_{\beta}\over 
4\nu{\cal R}}
{\partial\over \partial k_{\beta}}\delta \tilde{v}^{\alpha}_{T(0)}\left(t, 
\vec{k}\right)
-{1\over 2}U^{\alpha}C{\left(\vec{b}\wedge\vec{U}\right)_{\beta}\over 
4\nu{\cal R}}
\delta \tilde{v}^{\beta}_{T(0)}\left(t, \vec{k}\right)\ ,\\
\label{AA3}
&&{\partial\over \partial t}\delta \tilde{v}^{\alpha}_{T(2)}\left(t, 
\vec{k}\right) + \nu k^2\delta \tilde{v}^{\alpha}_{T(2)}\left(t,
\vec{k}\right) +{i\over 
2}\vec{k}\cdot\vec{U}\delta \tilde{v}^{\alpha}_{T(2)}\left(t, 
\vec{k}\right)\nonumber\\
&&={1\over 
2}\vec{k}\cdot\vec{U}C{\left(\vec{b}\wedge\vec{U}\right)_{\beta}\over 
4\nu{\cal R}}
{\partial\over \partial k_{\beta}}\delta \tilde{v}^{\alpha}_{T(1)}\left(t, 
\vec{k}\right)
-{1\over 2}U^{\alpha}C{\left(\vec{b}\wedge\vec{U}\right)_{\beta}\over 
4\nu{\cal R}}
\delta \tilde{v}^{\beta}_{T(1)}\left(t, \vec{k}\right)\ ,\\
&&...................................\nonumber
\ea
This system of equations yields the perturbative solution
\ba
\delta \tilde{v}^{\alpha}_T\left(t, \vec{k}\right) &&= e^{-\left(\nu
k^2+{i\over
2}\vec{U}\cdot\vec{k}\right)t}\Bigg\{F^{\alpha}_{(0)}\left(\vec{k}\right)+ 
F^{\alpha}_{(1)}\left(\vec{k}\right)\nonumber\\ &&+
C{\vec{k}\cdot\vec{U}\over 8\nu{\cal 
R}}\left[\left(\vec{b}\wedge\vec{U}\right)\cdot\vec{\nabla}_k
F^{\alpha}_{(0)}\left(\vec{k}\right) t -{U^{\alpha}\over 
\vec{k}\cdot\vec{U}}
\left(\vec{b}\wedge\vec{U}\right)\cdot\vec{F}_{(0)}\left(\vec{k}\right)t\right
.\nonumber\\
&&\left.- 
\left(\vec{b}\wedge\vec{U}\right)\cdot\vec{k}F^{\alpha}_{(0)}\left(\vec{k}
\right)\nu 
t^2\right]
+ O\left({1\over {\cal R}^2}\right)\Bigg\}\ 
\label{sol-0}
\ea
where the functions $F$'s are determined by the initial conditions: they
are found to be of $O(1)$ for any $k$.\\

The exponential term in front of (\ref{sol-0}) makes the 
perturbative solution vanish in the limit of large time $t$, provided
the perturbative series contained in the curly brackets does not diverge
faster in such a limit. 
This requirement can be translated into the following spectral condition
\be
{8\nu^2{\cal R}\over U^2}k^2>1\ .
\ee
This inequality indicates that the instability of solution (\ref{soluz})
may originate only from sufficiently small values of the wave--number
$k$.

\section{}
\label{B}
As shown in Section III the solution $\bar v^{\alpha}_T$ 
of the hydrodynamic operator in the action functional (\ref{act}) 
is defined for $t>0$. Accordingly, it breaks Galilean invariance,
thus giving rise to the well-known Doppler effect, i.e.
$k_0\rightarrow k_0 + {1\over 2}\vec{k}\cdot\vec{U}$.\\

Moreover, since in Section IV we evaluate the action functional
by applying a saddle--point expansion around $\bar v^{\alpha}_T$,
the approximated expression (\ref{FF}) contains a time integral
that has to be restricted to $t>0$ only. This amounts to assume that the
action should be identically zero for $t<0$. Accordingly, one cannot
exclude the possibility that a singularity in the time integral
may originate at $t=0$.

In this appendix we want to show that one can easily exclude the 
presence of any singularity by passing to a Fourier--transformed 
representation of the action functional (\ref{FF}): according to
a standard field-theoretic technique the addition of a small immaginary
part to the frequency appearing in the Fourier--transformed integral
allows one to control its regular behavior for $t \to 0^+$.

For the sake of clarity, we present this procedure only for two of the
terms appearing in (\ref{FF}). Actually, one can easily realize that
the procedure can be extended to all the terms: we just report the
final result, thus avoiding the writing of lengthy formulae.

Let us consider the term
\be
I_1 
= \int_0^{\infty} dt\int d^3x\int d^3y {\partial\over \partial t}
u^{\alpha}_T(t,\vec{x})F^{-1\alpha\beta}(|\vec{x}-\vec{y}|)
{\partial\over \partial t}u^{\beta}_T(t,\vec{y})
\label{GG1}
\ee
In principle, the integral in the time domain is ill--defined. We can
pass to Fourier--transformed variables and rewrite it as follows:

\be
I_1 
=-\int_{-\infty}^{+\infty}{dk_0\over
2\pi}\int_{-\infty}^{+\infty}{dq_0\over 2\pi}
\int {d^3k\over (2\pi)^3}
\int_0^{+\infty}dt\ e^{i(k_0+q_0)t}\tilde{u}^{\alpha}_T(k_0, \vec{k})
{k_0q_0\over F^{\alpha\beta}(k)}\tilde{u}^{\beta}_T(q_0, -\vec{k}).
\label{GG}
\ee
The time integral can be regularized by adding a small immaginary
part $i\epsilon$ to the frequency component and the integral
$I_1$ is transformed into 
\ba
I_1^{\prime} &&=-
\int {d^3k\over (2\pi)^3}
\int_{-\infty}^{+\infty}{dk_0\over
2\pi}\int_{-\infty}^{+\infty}{dq_0\over 2\pi}
\int_0^{+\infty}dt\ e^{i(k_0+q_0 + i\epsilon)t}\tilde{u}^{\alpha}_T(k_0,
\vec{k}) k_0q_0\tilde{u}^{\beta}_T(q_0, -\vec{k})\nonumber\\
&&= 
\int {d^3k\over (2\pi)^3}
\int_{-\infty}^{+\infty}{dk_0\over
2\pi}\int_{-\infty}^{+\infty}{dq_0\over 2\pi}{k_0q_0\over i(k_0+q_0 +
i\epsilon)}\tilde{u}^{\alpha}_T(k_0,\vec{k})\tilde{u}^{\beta}_T(q_0,
-\vec{k})\nonumber\\
&&= 
\int {d^3k\over (2\pi)^3}
\int_{-\infty}^{+\infty}{dk_0\over 2\pi}
\int_{-\infty}^{+\infty}{dq_0\over 2\pi}{k_0(q_0-k_0)\over i(q_0 +
i\epsilon)}\tilde{u}^{\alpha}_T(k_0,\vec{k})
\tilde{u}^{\beta}_T(q_0-k_0,-\vec{k}) \ .
\ea
By performing the limit $\epsilon \to 0^+$ one obtains
\ba
I_1^{\prime} &&= -
\int {d^3k\over (2\pi)^3}
i\int_{-\infty}^{+\infty}{dk_0\over 2\pi}\left[P
\int_{-\infty}^{+\infty}{dq_0\over 2\pi}{1\over
q_0}k_0(q_0-k_0)\tilde{u}^{\alpha}_T(k_0,\vec{k})
\tilde{u}^{\beta}_T(q_0-k_0,-\vec{k})\right.\nonumber\\
&&\left.-i\pi\int_{-\infty}^{+\infty}{dq_0\over
2\pi}\delta(q_0)k_0(q_0-k_0)\tilde{u}^{\alpha}_T(k_0,\vec{k})
\tilde{u}^{\beta}_T(q_0-k_0,-\vec{k})\right]\nonumber\\ 
&&= -i\int {d^3k\over (2\pi)^3}
\int_{-\infty}^{+\infty}{dk_0\over 2\pi}\left[{1\over 2\pi}
P\int_{-\infty}^{+\infty}dq_0{k_0(q_0-k_0)\over q_0}
\tilde{u}^{\alpha}_T(k_0,\vec{k})
\tilde{u}^{\beta}_T(q_0-k_0,-\vec{k})\right.\nonumber\\
&&\quad \left.+ {i\over 2} k_0^2\tilde{u}^{\alpha}_T(k_0,\vec{k})
\tilde{u}^{\beta}_T(-k_0,-\vec{k})\right]
\ea
In this equation $P$ denotes the principal value. The nontrivial part to be 
computed is contained in the square brackets. One has to consider that
the fluctuations 
$u^{\alpha}_T(t, \vec{x})$ become negligible for scales smaller than 
the Kolmogorov scale. Since they are defined for
$t>0$ and the time integral is singular in $t=0$, we have that 
its Fourier--transformed representation should exhibit a unique singularity 
at infinity, where it vanishes for $Im\ q_0<0$. One can write:
\ba
&&{1\over 2\pi}
P\int_{-\infty}^{+\infty}dq_0{k_0(q_0-k_0)\over q_0}
\tilde{u}^{\alpha}_T(k_0,\vec{k})
\tilde{u}^{\beta}_T(q_0-k_0,-\vec{k})\nonumber\\
&&= - {k_0^2\tilde{u}^{\alpha}_T(k_0,\vec{k})\over 2\pi}
P\int_{-\infty}^{+\infty}dq_0
{\tilde{u}^{\beta}_T(q_0-k_0,-\vec{k})\over q_0}\nonumber\\
&&= {i\over 2}k_0^2\tilde{u}^{\alpha}_T(k_0,\vec{k})
\tilde{u}^{\beta}_T(-k_0,-\vec{k})\ .
\ea
Making use of this result, one can easily conclude that (\ref{GG}) can
be written as follows:
\ba
I_1 = \int {dk_0d^3k\over (2\pi)^4}k_0^2\tilde{u}^{\alpha}_T(k_0,\vec{k})
F^{-1\alpha\beta}(k)\tilde{u}^{\beta}_T(-k_0,-\vec{k})\ .
\label{final}
\ea
Now, let us consider one of the terms of (\ref{FF}) which exhibits the
Doppler effect in its Fourier--transformed representation: 
\ba
I_2 &&= \int_0^{\infty} dt\int d^3x\int d^3y {\partial\over \partial t}
u^{\alpha}_T(t,\vec{x})F^{-1\alpha\beta}(|\vec{x}-\vec{y}|)
\bar{v}^{\lambda}_T(t,\vec{y})
\partial_{\lambda}u^{\beta}_T(t,\vec{y})\nonumber\\
&&= {U^{\lambda}\over 2}\int {d^3k\over (2\pi)^3}
\int_{-\infty}^{+\infty}{dk_0\over 2\pi}
\int_{-\infty}^{+\infty}{dq_0\over 2\pi}\left\{\int_0^{+\infty}dt\
e^{i(k_0+q_0 +
i\epsilon)t}\tilde{u}^{\alpha}_T(k_0,\vec{k}){k_0k_{\lambda}\over
F^{\alpha\beta}(k)}\tilde{u}^{\beta}_T(q_0,-\vec{k})\right.\nonumber\\
&&+ \int_0^{+\infty}dt\ e^{i(k_0+q_0 + i{U^2\over 4\nu{\cal R}^2})t}
{i\over 2}\tilde{u}^{\alpha}_T(k_0,\vec{k}){k_0\over
F^{\alpha\beta}(k)}\left[-\left(k_{\lambda}+
C{(\vec{b}\wedge\vec{U})_{\lambda}\over 4\nu {\cal R}}\right)
\tilde{u}^{\beta}_T\left(q_0, -\vec{k}-C{\vec{b}\wedge\vec{U}\over
4\nu {\cal R}}\right)\right.\nonumber\\
&&\left.\left.+ \left(k_{\lambda}-
C{(\vec{b}\wedge\vec{U})_{\lambda}\over 4\nu {\cal R}}\right)
\tilde{u}^{\beta}_T\left(q_0, -\vec{k}+C{\vec{b}\wedge\vec{U}\over
4\nu {\cal R}}\right)\right]\right\}\nonumber\\
&&= {1\over 2}\int {d^3k\over (2\pi)^3}
\int_{-\infty}^{+\infty}{dk_0\over 2\pi}
\int_{-\infty}^{+\infty}{dq_0\over 2\pi}\left\{{i\over k_0+q_0 +
i\epsilon}\tilde{u}^{\alpha}_T(k_0,\vec{k}){k_0(\vec{k}\cdot\vec{U})\over
F^{\alpha\beta}(k)}\tilde{u}^{\beta}_T(q_0,-\vec{k})\right.\nonumber\\
&&\left.-{1\over (k_0+q_0 + i{U^2\over 4\nu{\cal R}^2})}
\tilde{u}^{\alpha}_T(k_0,\vec{k}){k_0(\vec{k}\cdot\vec{U})\over
F^{\alpha\beta}(k)}C{(\vec{b}\wedge\vec{U})_{\lambda}\over 4\nu {\cal R}}
{\partial\over \partial k_{\lambda}}\tilde{u}^{\beta}_T(q_0,-\vec{k})
+O\left({1\over {\cal R}^2}\right)\right\}\ .
\ea
We expand the 
solution $ \bar{v}^{\lambda}_T$ up to first order in
powers of ${\cal R}^{-1}$ and we obtain the final expression:
\ba
I_2 = {1\over 2}\int {dk_0d^3k\over (2\pi)^4}
\tilde{u}^{\alpha}_T(k_0,\vec{k})
{k_0(\vec{k}\cdot\vec{U})\over F^{\alpha\beta}(k)}\left\{
\tilde{u}^{\beta}_T(-k_0,-\vec{k})
+iC{(\vec{b}\wedge\vec{U})_{\lambda}\over 8\nu {\cal R}}{\partial\over 
\partial k_{\lambda}}\tilde{u}^{\beta}_T(-k_0,-\vec{k})
+ O\left({1\over {\cal R}^2}\right)\right\}\ .\nonumber 
\ea
As in the previous case, one can regularize the integral in $t=0$
by performing the limit $\epsilon \to 0^+$.
By applying this procedure to all of the remaining terms in (\ref{FF})
one arrives at the final expression (\ref{act1}).

\section{}
\label{C}

In this Appendix we sketch the calculation of the eigenvalues of the
matrix $M^{\beta}_{\zeta}(\hat{p})$ defined in (\ref{matr1}).
In fact, the perturbative expansion of
the solution (\ref{soluz}) in powers of$1\over \cal R$
induces an analogous expansion for this matrix. Formally, one can
write
\be
M = M_{(0)} + M_{(1)} + ...
\label{Mexp}
\ee
where
\ba
M^{\alpha}_{(0)\ \beta} &&= \delta^{\alpha}_{\ \beta}\left[i\left(p_0
+ {1\over 2}\vec{p}\cdot \vec{U}\right)
+\nu p^2 \right]\ ,\nonumber\\
M^{\alpha}_{(1)\ \beta} &&= -\delta^{\alpha}_{\ \beta}{C\over 
4}\vec{p}\cdot\vec{U}
{\left(\vec{b}\wedge\vec{U}\right)^{\gamma}\over 4\nu{\cal 
R}}\partial_{p_{\gamma}}
-{C\over 4}U^{\alpha}{\left(\vec{b}\wedge\vec{U}\right)_{\beta}\over 
4\nu{\cal R}}\ .
\ea
The matrix $M^{\beta}_{\zeta}(\hat{p})$
acts on the two-dimensional space of the transverse functions and on
the one--dimensional space of the longitudinal functions.
Only the transverse degrees of freedom are physically
relevant.\\

A complete orthonormal basis in $R^3$ is given by the vectors
\ba
\Pi_1^{\alpha} &&= {\left(\vec{b}\wedge\vec{p}\right)^{\alpha}\over 
\sqrt{f(p)}}\ ,
\nonumber\\
\Pi_2^{\alpha} &&= {g(p)\left(\vec{b}\wedge\vec{p}\right)^{\alpha}
- f(p)\left(\vec{U}\wedge\vec{p}\right)^{\alpha}\over
\sqrt{f(p)}\sqrt{f(p)h(p)-g^2(p)}},\nonumber\\
\Pi_3^{\alpha} &&= {p^{\alpha}\over p}\ ,
\ea
where 
\ba
f(p) = b^2p^2-(\vec{b}\cdot\vec{p})^2, \quad g(p) = (\vec{b}\cdot\vec{U})p^2
- (\vec{b}\cdot\vec{p})(\vec{U}\cdot\vec{p}), \quad h(p) =
U^2p^2-(\vec{U}\cdot\vec{p})^2\ .
\ea
$\Pi_1^{\alpha}$ and $\Pi_2^{\alpha}$ span the transverse
subspace, while
$\Pi_3^{\alpha}$ spans the longitudinal one. 
In analogy with (\ref{Mexp}), also the eigenvalues of 
$M^{\beta}_{\zeta}(\hat{p})$ can be represented by a perturbative expansion
in powers of$1\over \cal R$, namely
as
\be
\lambda^a = \lambda^a_{(0)} + \lambda^a_{(1)} + ...\quad where\quad a=1, 
2, 3\ .
\ee
The zero-order eigenvalues $\lambda^a_{(0)}$ are degenerate and 
have the form
\be
\lambda^a_{(0)} =\left(i\left(p_0 + {1\over 2}\vec{p}\cdot 
\vec{U}\right) +\nu
p^2\right)\ .
\ee
The evaluation of the first order corrections $\lambda^a_{(1)}$ requires the
diagonalization of the matrix with elements $M_{(1)ij}=\left(\Pi_i,
M_{(1)}\Pi_j\right)$,  ($i,j=1, 2, 3$). After some simple but lengthy
calculations one finds
\ba
\lambda^1_{(1)} &&= {1\over 2}\left(M_{(1)11}+ M_{(1)22}
- \sqrt{\left(M_{(1)11}+ M_{(1)22}\right)^2 + 4M_{(1)21}M_{(1)12}}\right)\
,\nonumber\\
\lambda^2_{(1)} &&= {1\over 2}\left(M_{(1)11}+ M_{(1)22}
+ \sqrt{\left(M_{(1)11}+ M_{(1)22}\right)^2 + 4M_{(1)21}M_{(1)12}}\right)\
,\nonumber\\
\lambda^3_{(1)} &&= M_{(1)33}\ ,
\ea
with
\ba
M_{(1)11} &&={C\over 16\nu{\cal R}}{\left(\vec{b}\wedge\vec{U}\right)
\cdot\vec{p}\over f(p)}w(p)\ ,\nonumber\\
M_{(1)22} &&= - {C\over 16\nu{\cal 
R}}{\left(\left(\vec{b}\wedge\vec{U}\right)
\cdot\vec{p}\right)\left(\vec{b}\cdot\vec{p}\right)g(p)\over
f(p)\left(f(p)h(p)-g^2(p)\right)}\left[\left(\vec{p}\cdot\vec{U}\right)w(p)
+ (\vec{b}\cdot\vec{U})g(p) -U^2f(p)\right],\nonumber\\
M_{(1)12} &&=- {C\over 16\nu{\cal
R}}{\left(\left(\vec{b}\wedge\vec{U}\right)\cdot\vec{p}\right)
\left(\vec{b}\cdot\vec{p}\right)\over
f(p)\sqrt{f(p)h(p)-g^2(p)}}\left[\left(\vec{b}\cdot\vec{U}\right)g(p)
+2\left(\vec{p}\cdot\vec{U}\right)w(p)
-U^2f(p)\right],\nonumber\\
M_{(1)21} &&= -{C\over 16\nu{\cal 
R}}{\left(\left(\vec{b}\wedge\vec{U}\right)
\cdot\vec{p}\right)\over f(p)\sqrt{f(p)h(p)-g^2(p)}}
\left(\vec{b}\cdot\vec{U}\right)\left[\left(\vec{b}\cdot\vec{p}\right)g(p)
- \left(\vec{p}\cdot\vec{U}\right)f(p)\right],\nonumber\\
M_{(1)33} &&= -{C\over 16\nu{\cal R}}{\left(\vec{p}\cdot\vec{U}\right)\over
p^2}\left(\left(\vec{b}\wedge\vec{U}\right)\cdot\vec{p}\right)\ ,
\ea
where we have introduced the further definition:
\be
w(p) = b^2(\vec{p}\cdot\vec{U}) - 
(\vec{b}\cdot\vec{p})(\vec{b}\cdot\vec{U})\ .
\ee
Without prejudice of generality, we can  
specify the geometrical structure of the flow. For the
sake of simplicity, we assume that the vector $\vec{r}$ (i.e. the 
Fourier--conjugated variable of $\vec{p}$) corresponds to the polar 
axis and that the vector
$\vec{b}$ is orthogonal to both
$\vec{r}$ and $\vec{U}$. With this assumption the two physically 
relevant first-order corrections to the eigenvalues are
\ba
\lambda^1_{(1)} &&= 0\ ,\nonumber\\
\lambda^2_{(1)} &&= {U^2\over 16\nu {\cal R}}\left\{
\sin\theta_U\cos\theta_U\left[\cos^2\phi_U + 
\cos\left(2(\phi_U-\phi)\right)\right]
\sin^2\theta
\right.\nonumber\\
&&\left.+\cos^2\theta_U\sin 2\theta\cos(\phi_U -\phi)\right\}\ .
\label{B1}
\ea
Since $\lambda_{(i)}^{3}$ is associated to the longitudinal
part, it does not play any role in our calculations.

\section{}
\label{D}
In this appendix we aim at reporting the main calculations
needed for obtaining an explicit expression for (\ref{S21}).
According to the perturbative approach discussed in detail
in Appendix \ref{C}, $S_2(r)$ can be written as follows:
\be
S_2(r)\sim -2\int {dp_0d^3p\over (2\pi)^4}\left(e^{i\vec{p}\cdot\vec{r}}
- 1\right)\sum_{\alpha=1}^2{F(p)\over \left(p_0+{1\over 
2}\vec{p}\cdot\vec{U}\right)^2
+ \left(\nu p^2 +
\lambda^{\alpha}_{(1)}
(\vec{p},\vec{U}, \vec{b}) 
\right)^2}\ .
\ee
The eigenvalues $ \lambda^{\alpha}_{(1)} $ which appear in this
equation have been computed up to first order of the perturbative
expansion in ${\cal R}^{-1}$. Notice that the sum is restricted
to the first two eigenvalues ($\alpha = 1,2$), which correspond to the
transverse components of the velocity field. Actually, the third
eigenvalue, corresponding to the longitudinal components of the velocity
field, is ineffective for our calculations.

Explicit integration over $p_0$ yields
\be
{S}_2(r) \sim -\int {d^3p\over (2\pi)^3}{e^{i\vec{p}\cdot\vec{r}} - 1\over
\nu}\sum_{\alpha=1}^2 {F(p)\over p^2 +{1\over 
\nu}\lambda^{\alpha}_{(1)}(\vec{p},\vec{U}, \vec{b})+
...}\
\label{S22}
\ee
With the particular choice performed in Appendix \ref{C} for the
geometrical structure of the flow, $S_2(r)$ can be expressed 
as the sum of two terms: the first one is associated with the null eigenvalue 
$\lambda_{(1)}^1$, while the second one depends on the nonzero eigenvalue
$\lambda_{(1)}^2$. Namely,
\begin{equation}
S_2(r) =-{1\over \nu}\left(I_1(r) +I_2(r)\right) 
\end{equation}
By considering the explicit expressions of the statistical function $F(p)$ and 
of the eigenvalues $\lambda^{\alpha}_{(1)}$ (see eq.(\ref{B1})~), one has
\ba
\label{C2}
I_1(r) &&= D_0L^3\int{d^3p\over (2\pi)^3}\left(e^{i\vec{p}\cdot\vec{r}}
- 1\right) {(Lp)^se^{-(Lp)^2}\over p^2}\ , \\
I_2(r) &&= D_0L^3\int{d^3p\over (2\pi)^3}
{\left(e^{i\vec{p}\cdot\vec{r}}
- 1\right)(Lp)^se^{-(Lp)^2}\over p^2+ {U^2\over 16\nu^2 {\cal R}}
\left[2\sin\theta_U\cos\theta_U\sin^2\theta\cos^2\phi
+\cos^2\theta_U\sin 2\theta\cos \phi\right]}\ .\nonumber\\
\label{C3}
\ea
In the r.h.s. of this equation we have also exploited 
translational invariance for applying the transformation
$(\phi_U-\phi)\rightarrow -\phi$. 
The analytic calculation of (\ref{C2}) is obtained by a standard
procedure:
\ba
I_1(r) &&= D_0L^3\int{d^3p\over (2\pi)^3}\left(e^{i\vec{p}\cdot\vec{r}}
- 1\right) {(Lp)^se^{-(Lp)^2}\over p^2}\nonumber\\
&&= {D_0L^2\over 2\pi^2}\sum_{n=1}^{\infty}{(-1)^n\over
(2n)!(2n+1)}\left({r\over L}\right)^{2n}\int_0^{\infty} d\zeta\
\zeta^{s+2n}e^{-\zeta^2}\nonumber\\
&&= {D_0\over (2\pi)^2}r^2\sum_{n=0}^{\infty}(-1)^{n+1}
{\Gamma\left({s+3+2n\over 2}\right)\over
\Gamma\left(2n+4\right)}\left({r\over L}\right)^{2n}\ .
\ea
For what concerns $I_2(r)$, we first perform the integration over
the variable $\phi$, namely:
\be
I_2(r) = D_0L^3\int_0^{\infty}{p^2dp\over (2\pi)^3}(Lp)^se^{-(Lp)^2}
\int_{-1}^{+1} d(\cos\theta)\left(e^{ipr\cos\theta} - 1\right)I_0
\label{C5}
\ee
where
\ba
I_0 &&=\int_0^{2\pi}{d\phi\over p^2+ {U^2\over 16\nu^2 {\cal R}}
\left[2\sin\theta_U\cos\theta_U\sin^2\theta\cos^2\phi
+\cos^2\theta_U\sin 2\theta\cos \phi\right]}\nonumber\\
&&=-i{32\nu^2{\cal R}\over U^2}\int_{\gamma}{zdz\over 
az^4+bz^3+cz^2+bz+a}\ ,
\label{C6}
\ea
with  $z=e^{i\phi}$ and the integration is on the unit circle
$\gamma$. The coefficients $a, b, c$ are given by
\ba
&&a = \sin\theta_U\cos\theta_U\sin^2\theta\ ,\quad
b=2\cos^2\theta_U\sin\theta\cos\theta\nonumber\\
&&c = {32\nu^2{\cal R}\over U^2}p^2 +
2\sin\theta_U\cos\theta_U\sin^2\theta\ .
\label{C7}
\ea
The evaluation of the integral (\ref{C6}) requires the knowledge of
the root of a fourth--order
algebrical equation. By exploiting the Euler method \cite{EUL} we end up
with the expression
\ba
z_i&&=z_i\left(x, {8\nu^2{\cal R}\over U^2}p^2; \Sigma, 
\Xi\right)\nonumber\\
&&={x^{1\over 3}\over \left(1-x^2\right)^{1\over 6}}\left[\sum_{l=0}^2s_{il}
F_l\left(x, {8\nu^2{\cal R}\over U^2}p^2; \Sigma, \Xi\right)
+ {1\over 2}\Sigma{x^{2\over 3}\over \left(1-x^2\right)^{1\over
3}}\right]
\quad \quad i=1, 2, 3, 4\ .\nonumber
\ea
The following definition has been adopted:
\ba
F_l&&=F_l\left(x, {8\nu^2{\cal R}\over U^2}p^2; \Sigma, 
\Xi\right)\nonumber\\
&&=
\left\{{\Sigma^{2\over 3}\over 12}\left[ {81\over 4}\Sigma^4 {x^4\over
\left(1-x^2\right)^2} + {81\over 2}\Sigma^2{x^2\over \left(1-x^2\right)}
 - 90\right.\right.\nonumber\\
&&-{64\over \Sigma^2}{1-x^2\over x^2} + {8\nu^2{\cal R}\over U^2}p^2
\left(189{\Sigma^2\over \Xi}{x^2\over \left(1-x^2\right)^2} + {382\over
\Xi\left(1-x^2\right)} -120{\Sigma^2\over \Xi\ x^2}\right)\nonumber\\
&&\left.+\left({8\nu^2{\cal R}\over U^2}p^2\right)^2
\left({504\over \Xi^2\left(1-x^2\right)^2} + {47\ \Sigma^2\over \Xi^2\
x^2\left(1-x^2\right)}\right) +
\left({8\nu^2{\cal R}\over U^2}p^2\right)^3{32\ \Sigma^2\over \Xi^3\ x^2
\left(1-x^2\right)^2}\right]^{1\over 3}\nonumber\\  
&&\times\left(\epsilon^l\left[1 + \left(1 -4\times 27\ h\right)^{1\over
2}\right]^{1\over 3}+\epsilon^{l-3}\left[1 - \left(1 -4\times 27\ 
h\right)^{1\over
2}\right]^{1\over 3}\right) + {1\over 2}{\Sigma^{4\over 3}x^{4\over 3}\over
\left(1-x^2\right)^{4\over 6}}\nonumber\\
&&\left.+{1\over 3}
{\left(1-x^2\right)^{1\over 3}\over \Sigma^{2\over 3}x^{2\over 3}} + 
{8\nu^2{\cal
R}\over U^2}p^2{2\over 3\Sigma^{2\over 3}\Xi\ x^{2\over 
3}\left(1-x^2\right)^{4\over
6}}\right\}^{1\over 2}\ ,
\label{C9}
\ea
with
\ba
\label{C10}
&&x=\cos\theta\ ,\quad
\Xi = \sin\theta_U\cos\theta_U\ ,\quad \Sigma =\cot\theta_U\ ,\\
\label{C11}
&&s_{il}\Leftrightarrow
\left(\begin{array}{ccc}
1 & 1 & 1\\
1 & -1 & -1\\
-1 & 1 & -1\\
-1 & -1 & -1
\end{array}
\right)\ .
\ea
Here $\epsilon$ is the cubic root of unit: $\epsilon = {-1+ i 
\sqrt{3}\over 2}$.
The explicit expression of the function $h$ follows:
\ba
h &&=\left[16 +30\
\Sigma^2{x^2\over 1-x^2} + {111\over 4}\Sigma^4{x^4\over(1-x^2)^2}
+16{8\nu^2{\cal R}\over U^2}{p^2\over
\Xi}{17\Sigma^2x^2+3(1-x^2)\over (1-x^2)^2}\right.\nonumber\\
&&\left.+48\left({8\nu^2{\cal R}\over U^2}{p^2\over \Xi}\right)^2{1\over
(1-x^2)^2}\right]^3\times\left[
-128 +81\ \Sigma^4{x^4\over (1-x^2)^2} + {81\over 2}\Sigma^6{x^6\over 
(1-x^2)^3}
\right.\nonumber\\
&&-180\Sigma^2{x^2\over 1-x^2}+{8\nu^2{\cal R}\over U^2}{p^2\over \Xi}
\left(378{\Sigma^4x^4\over (1-x^2)^3}+764{\Sigma^2x^2\over
(1-x^2)^2}-240{1\over 1-x^2}\right)\nonumber\\
&&\left.+\left({8\nu^2{\cal R}\over U^2}{p^2\over \Xi}\right)^2
\left(1008{\Sigma^2x^2\over (1-x^2)^3}+94{1\over (1-x^2)^2}\right)
+64\left({8\nu^2{\cal R}\over U^2}{p^2\over \Xi}\right)^3{1\over
(1-x^2)^3}\right]^{-2}\ .
\label{C12}
\ea
Only the roots $z_1$ and $z_2$ are included into the unit circle, 
therefore
(\ref{C5}) becomes
\ba
I_2(r) &&= D_0L^3{32\nu^2{\cal R}\over U^2}\int_0^{\infty}
{p^2dp\over  (2\pi)^2}(Lp)^se^{-(Lp)^2}\int_{-1}^1dx\left(e^{iprx} -
1\right)\nonumber\\
&&\times\sum_{l=1,2}{\left(1-x^2\right)^{1\over 3}\over x^{2\over 3}}
{\left[\sum_{m=0}^2s_{lm}
F_m\left(x, {8\nu^2{\cal R}\over U^2}p^2; \Sigma, \Xi\right)
+ {1\over 2}\Sigma{x^{2\over 3}\over \left(1-x^2\right)^{1\over
3}}\right]\over
\prod_{i\not= l}\left(\sum_{k=0}^2\left(s_{lk}-s_{ik}\right)
F_k\left(x, {8\nu^2{\cal R}\over U^2}p^2; \Sigma, \Xi\right)\right)}\ .
\label{C13}
\ea
As we have already observed in Section V, only the values of the
variable $p$ around
$\bar{p}= {1\over L}\sqrt{s+2\over 2}$ give a significant contribution
to the integral in (\ref{C13}). We observe that ${8\nu^2{\cal R}\over
U^2}\bar{p}^2\rightarrow{4(s+2)\over {\cal R}}$ and the
stability condition (\ref{cond1}) imposes:
\be
1<{\cal R}<4(s+2)\ .
\label{C14}
\ee
The evaluation of the leading terms is then possible by performing an
expansion in the parameter
${U^2\over 8\nu^2{\cal R}}p^{-2}\rightarrow {{\cal R}\over
8}\zeta^{-2}$ that,
by virtue of (\ref{C14}), is smaller than unit if
$\zeta<\sqrt{s+2\over 2}$.\\
 For
$\zeta>\sqrt{s+2\over 2}$ the contribution
 to the integral rapidly vanishes. For
$1<{\cal R}\ll 4(s+2)$ we obtain
\ba
\bar{S}_2(r) &&= -{1\over \nu}\left(I_1(r)+I_2(r)\right)\nonumber\\
&&\sim -{D_0\over
(2\pi)^2\nu}r^2\sum_{n=0}^{\infty}\left\{(-1)^{n+1}\Gamma\left({s+2n+3\over 
2}\right)
\left[{1+\Xi\over \Gamma\left(2n + 4\right)}
-{2^{13\over 3}\Xi\over \Sigma^{2\over 3}}{2n+4\over
\Gamma (2n+6)}\right]\left({r\over L}\right)^{2n}\right\}\ .\nonumber\\
\ea
By extending the validity of our calculations to ${\cal R}>4(s+2)$, we have
${8\nu^2{\cal R}\over U^2}p^2\rightarrow{8\over {\cal R}}\zeta^2<1$ for
$\zeta<\sqrt{s+2\over 2}$. As in the previous case,
we expand (\ref{C13})
in power of the parameter ${8\over {\cal R}}\zeta^2<1$ and we obtain:
\ba
I_2(r) &&\sim \Xi D_0L^2\int_0^{\infty}{d\zeta\over  
(2\pi)^2}\zeta^se^{-\zeta^2}
\int_{-1}^1dx\left(e^{i\zeta {r\over L}x} - 1\right)
\left\{{1 +{8\over {\cal R}}\zeta^2 +...\over 2}\right.\nonumber\\
&&+ {8\over {\cal R}\Xi}\sum_{l=1,2}{\left(1-x^2\right)^{1\over 3}\over 
x^{2\over 3}}
\left({\left[\sum_{m=0}^2s_{lm}
F_m\left(x, 0; \Sigma, \Xi\right) \right]\over
\prod_{i\not= l}\left(\sum_{k=0}^2\left(s_{lk}-s_{ik}\right)F_k
\left(x, 0; \Sigma, \Xi\right)\right)}\right.\nonumber\\
&&\left.\left.+ {8\over {\cal R}}\zeta^2\left.{\partial\over \partial y}
{\left[\sum_{m=0}^2s_{lm}
F_m\left(x, y; \Sigma, \Xi\right) \right]\over
\prod_{i\not= l}\left(\sum_{k=0}^2\left(s_{lk}-s_{ik}\right)F_k
\left(x, y; \Sigma, \Xi\right)\right)}\right|_{y=0} + ...\right)\right\}\ .
\label{C16}
\ea
Two different terms, $I_2^A(r)+I_2^B(r)= I_2(r)$,
can be identified in (\ref{C16}). The
evaluation of the first term is straightforward:
\ba
I_2^A(r) &&\sim {\Xi D_0L^2\over 2}\int_0^{\infty}
{d\zeta\over  (2\pi)^2}\zeta^se^{-\zeta^2}\int_{-1}^1dx\left(e^{i\zeta 
{r\over L}x} -
1\right)
\left(1 +{8\over {\cal R}}\zeta^2 + ...\right)\nonumber\\
&&= {\Xi D_0\over 2(2\pi)^2}r^2\sum_{n=0}^{\infty}{(-1)^{n+1}\over 
\Gamma(2n+4)}
\left(\Gamma\left({s+2+2n\over 2}\right)+ {8\over {\cal R}}
\Gamma\left({s+5+2n\over 2}\right)\right)\left({r\over 
L}\right)^{2n}\nonumber\\
&&\times\left(1 +{4(s+2)\over {\cal R}} + ...\right)\ .
\label{C17}
\ea
The evaluation of the second term is more cumbersome.
The leading term can be recasted in the form:
\ba
I_2^B(r) &&\sim {8D_0L^2\over {\cal R}}\int_0^{\infty}{d\zeta\over  
(2\pi)^2}\zeta^s
e^{-\zeta^2}\int_{-1}^1dx\left(e^{i\zeta {r\over L}x} -
1\right){\left(1-x^2\right)^{1\over 3}\over x^{2\over
3}}\sum_{n=0}^{\infty}A_n(\Sigma)x^{2n}\ .
\label{B2}
\ea
The coefficients $A_i$ are $\Sigma$-dependent numerical constants. The 
first two of them are given by the expressions
\ba
A_0(\Sigma) &&= {1\over 16\sqrt{3}\left(1-\sin{\pi\over 6}\right)
\cos\left({1\over 3}\tan^{-1}\sqrt{26}\right)}\ ,\nonumber\\
A_1(\Sigma) &&= -{65\sin\left({2\over 3}\tan^{-1}\sqrt{26}\right)\over
512\sqrt{26}\cos^2\left({1\over 3}\tan^{-1}\sqrt{26}\right)}\Sigma^2\ ,...
\ea 
The exact form of these coefficients is however irrelevant for our 
analysis.
Some tedious standard calculations yield:
\ba
I^B_2(r) &&= {D_0{\cal R}^{1\over 3}\over \pi\Gamma\left({2\over 3}\right)}
\left({\nu\over U}\right)^{4\over 3}r^{2\over 3}
\sum_{n=0}^{\infty}C_n(\Sigma)\Gamma\left({3s+3n+5\over 6}\right)
\left({r\over L}\right)^n\ ,
\ea
where the coefficients $C_n(\Sigma)$ depend on the constants 
$A_i$. For $n=0$ one has
\be
C_0(\Sigma) = {54\sqrt{3}-74\over 27\sqrt{3}}A_0 + {128\over
9\sqrt{3}}A_1(\Sigma)\ .
\ee
The comparison between $I_2^B(r)$ and $I_2^A(r)$ 
indicate that a crossover between the corresponding scaling behaviors
occurs at
\be
r\sim \left|2\times 8.328 \sqrt{\pi}{0.0336 - 0.1127\cot^2\theta_U\over 2+
\sin\theta_U\cos\theta_U}\right|^{3\over 4} {\cal R}^{-{3\over 4}}L\ .
\ee
For the perturbative expansion in $1\over {\cal R}$ to be meaningful, 
the parameter $\theta_U$ must have a value close to
$\pi\over 2$. This implies:
$$
r\sim F{\cal R}^{-{3\over 4}}L, \quad with\quad F\sim 0.6\ .
$$
\end{appendix}

\end{document}